\documentclass[a4paper]{article}
\usepackage{ASVspoof}
\usepackage{epsfig,amssymb,amsmath}
\usepackage{flushend}

\usepackage{array}
\usepackage[table]{xcolor}
\usepackage{hhline}
\usepackage{makecell}
\usepackage{multirow,booktabs}
\ninept

\title{ASASVIcomtech: The Vicomtech-UGR Speech Deepfake Detection and SASV Systems for the ASVspoof5 Challenge} %ASASVIcomtech: The Vicomtech-UGR speech deepfake detection and SASV systems to the ASVspoof5 Challenge
\makeatletter
\def\name#1{\gdef\@name{#1\\}}
\makeatother
\name{{\em Juan M. Martín-Doñas$^1$, Eros Roselló$^2$, Angel M. Gomez$^2$,}\\
      {\em Aitor Álvarez$^1$, Iván López-Espejo$^2$ and Antonio M. Peinado$^2$}}
\address{$^1$Fundación Vicomtech, Basque Research and Technology Alliance (BRTA), San Sebastián, Spain\\
  $^2$Dept. of Signal Theory, Telematics and Communications -- CITIC, University of Granada, Spain \\
{\small \tt \{jmmartin,aalvarez\}@vicomtech.org, \{erosrosello,amgg,iloes,amp\}@ugr.es} }
\begin{document}
\maketitle

\begin{abstract}
This paper presents the work carried out by the ASASVIcomtech team, made up of researchers from Vicomtech and University of Granada, for the ASVspoof5 Challenge. The team has participated in both Track 1 (speech deepfake detection) and Track 2 (spoofing-aware speaker verification). This work started with an analysis of the challenge available data, which was regarded as an essential step to avoid later potential biases of the trained models, and whose main conclusions are presented here. With respect to the proposed approaches, a closed-condition system employing a deep complex convolutional recurrent architecture was developed for Track 1, although, unfortunately, no noteworthy results were achieved. On the other hand, different possibilities of open-condition systems, based on leveraging self-supervised models, augmented training data from previous challenges, and novel vocoders, were explored for both tracks, finally achieving very competitive results with an ensemble system.
\end{abstract}

\section{Introduction}
\label{sec:intro}

%%% INTRODUCCIÓN DE ANTONIO %%%
%Recent advances in deep learning techniques for the generation of synthetic speech mimicking the voice of a certain speaker poses a challenging threat to our society \cite{Masood23}. 
%For example, these technologies can be used in authentication systems based on automatic speaker verification (ASV) to supplant a given user by means of spoofed speech \cite{Wu14}. Also, they can be used to forge voice deepfakes with the aim of blackmailing or vilifying somebody \cite{Yamagishi21}. The main technologies behind these threats are text-to-speech synthesis (TTS) and voice conversion (VC).

Recent advances in deep learning techniques for the generation of synthetic speech mimicking the voice of a certain speaker poses a challenging threat to our society \cite{Masood23}. These generative models, driven by text-to-speech synthesis (TTS) and voice conversion (VC) technology, have legitimate applications~\cite{Gonzalez20} but can be misused for malicious purposes, such as forging voice deepfakes with the aim of blackmailing or vilifying somebody \cite{Yamagishi21}. Moreover, these models can be used in authentication systems based on automatic speaker verification (ASV) to supplant a given user by means of spoofed speech \cite{Wu14}.

The scientific community has responded to this situation by a series of challenges which have set up unified development frameworks, thus allowing the establishment of benchmarks and making it possible quick comparisons among different countermeasure (CM) systems. Examples are the ASVspoof challenge series 2015--21 \cite{ASVspoof}, the Audio Deepfake Detection (ADD) challenges 2022--23 \cite{Yi22,Yi23}, and the Spoofing-Aware Speaker Verification (SASV) Challenge 2022 \cite{SASV22}.

This year, the fifth ASVspoof challenge (ASVspoof5) was launched and recently wrapped up \cite{Wang2024_ASVspoof5}. This time, the challenge was organized in two tracks: \emph{1)} standalone spoofing and speech deepfake detection (non-ASV), and \emph{2)} spoofing-aware automatic speaker verification (SASV), where participants could develop their own joint ASV-CM systems. For both tracks, two conditions were considered: closed (developments restricted to ASVspoof5 training data), and open (external data and pre-trained models were also allowed).

This paper presents the work carried out by the ASASVIcomtech team, comprised of researchers from Vicomtech and the University of Granada, for the ASVspoof5 Challenge. The team has participated in both Track 1 (closed and open conditions) and Track 2 (open condition only).

For the closed condition of Track 1, we applied a deep complex convolutional recurrent network (DCCRN) fed with full-spectrum features derived from the short-time Fourier transform (STFT). To adapt the DCCRN to this task, we only utilized the CNN encoder and recurrent LSTM layers, omitting the decoder part. Thus, the last LSTM state is projected onto an embedding and then passed through a softmax layer for classification.

For the open condition of Tracks 1 and 2, the team has proposed an ensemble system based on two self-supervised models (Wav2Vec2-Large \cite{Baevski20} and WavLM-Base \cite{Chen22}) as deep feature extractors for the CM part. 
%These deep features are then fed to fine-tuned downstream classifiers to compute the CM scores. 
Downstream classifiers are then fine-tuned to compute the CM scores from these deep features.
To obtain the ASV scores required for Track 2, we have considered the TitaNet-Large ASV model \cite{TitaNet} for embedding extraction and cosine scoring. The final SASV scores are achieved from the calibrated log-likelihood ratio (LLR) ASV and CM scores via non-linear fusion.

The rest of this paper is organized as follows. Section \ref{sec:asvspoof5} outlines the preliminary analysis performed on the training and validation data. Then, we describe the corresponding systems and challenge results for closed and open conditions in Sections \ref{sec:closed} and \ref{sec:open}, respectively. Finally, the paper concludes with a summary of the work in Section \ref{sec:conclusion}.
%%%%%%%%%%%%%%%%%%%%%%%%%%%%%

\section{ASVspoof5 dataset analysis} \label{sec:asvspoof5}

Before designing our systems for the ASVspoof5 Challenge, we conducted a preliminary analysis of the new training and development datasets. This analysis guided several key decisions in our design process, making it pertinent to include our findings in this paper. In this section, we will examine the database provided for the challenge, focusing on data balance, utterance duration, delays, and speech quality distributions.

\subsection{Balancing}

\begin{figure}[t]
\centering
\includegraphics[width=0.49\textwidth]{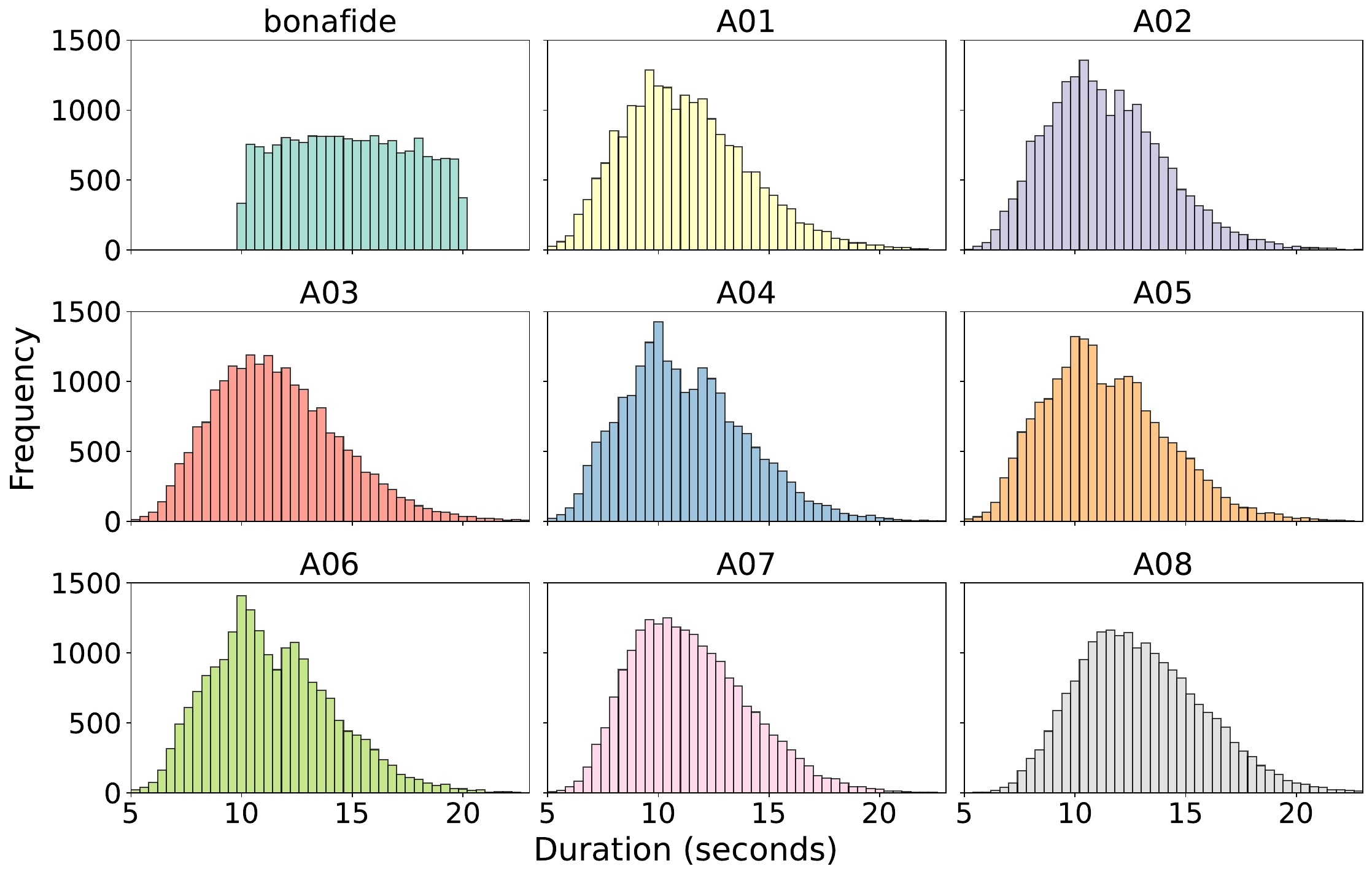}
%\caption{Training set utterance duration histograms.}\label{fig:Histogram_by_Attacks_train}
\caption{\it Histograms of utterance duration from the training set.}\label{fig:Histogram_by_Attacks_train}
%\vspace{0.35cm}
\end{figure}
%Training utterance duration histograms for the different attack types.

{\bf Training:} The ASVspoof5 training partition contains a total of 182,357 utterances, with 18,797 labeled as bonafide and the remaining samples, as spoofed. Each spoofing attack type (A01-A08) includes 20,445 samples. Therefore, approximately 10\% of the training data is bonafide. The dataset is gender-balanced, with roughly 50\% of the utterances being male-voiced (92,236 utterances) and 50\% female-voiced (90,121 utterances).

\noindent
{\bf Development:} The ASVspoof5 development partition includes 140,950 utterances, with 31,334 being bonafide and the rest, spoofed. Each spoofing attack type (A09-A16) comprises 13,702 samples. In contrast to the training set, approximately 22\% of the development data is bonafide. This differs from previous ASVspoof editions, such as that of 2019, where the training and validation datasets had a similar percentage of bonafide audio samples. Like the training set, the development set is also gender-balanced, with about 50\% male utterances (71,863 utterances) and 50\% female utterances (69,087 utterances).

\subsection{Duration of the utterances}

\begin{figure}[t]
\includegraphics[width=0.49\textwidth]{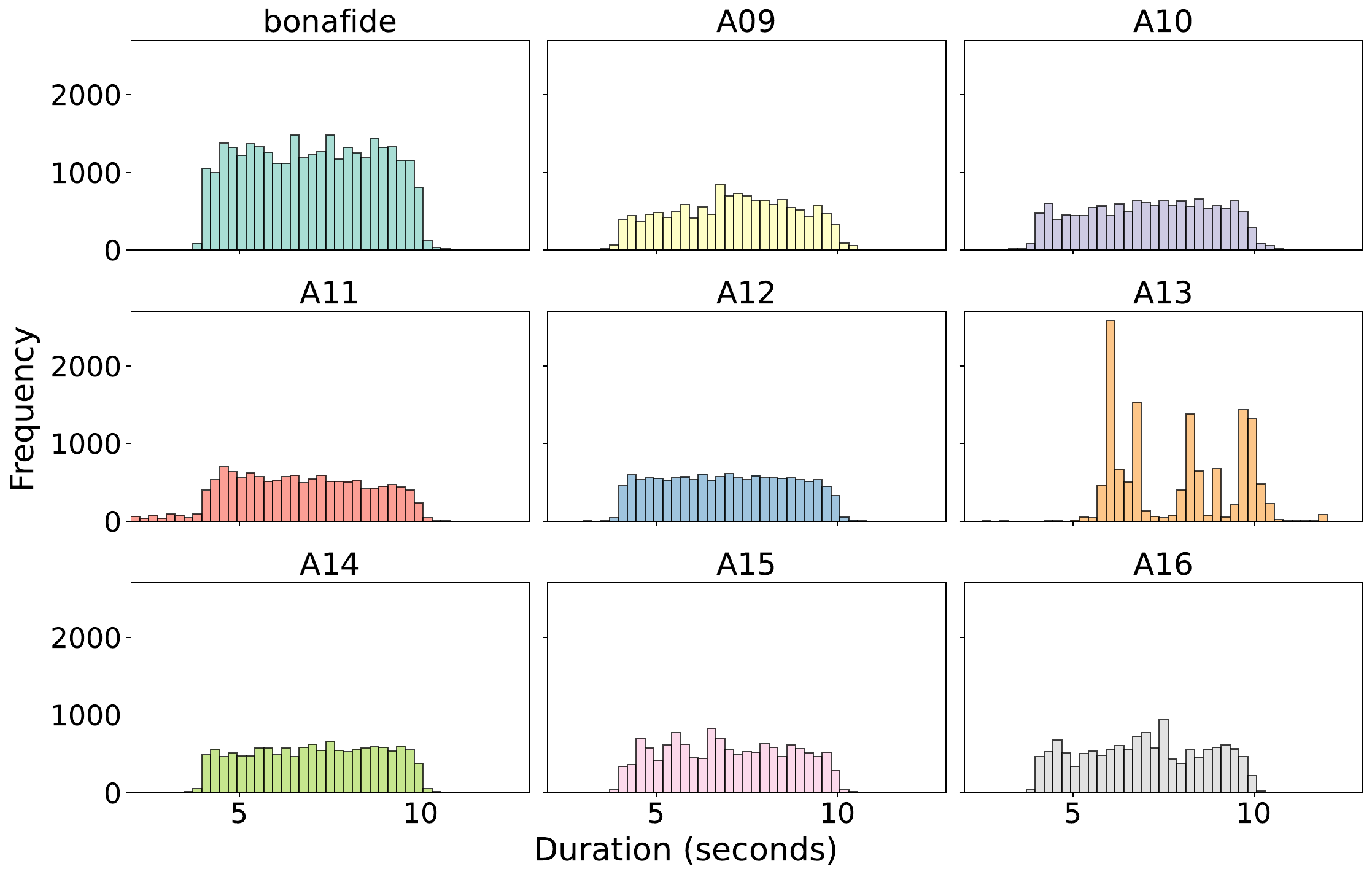}
%\caption{Development set utterance duration histograms.}\label{fig:Histogram_by_Attacks_dev}
\caption{\it Histograms of utterance duration from the development set.}\label{fig:Histogram_by_Attacks_dev}
\end{figure}

{\bf Training:} The average duration of utterances in the training dataset is 11.92 s, with a standard deviation of 2.99 s. This indicates that the audios contain significantly more information compared to those of the 2019 ASVspoof Challenge, where the training and validation dataset utterances had an average duration of around 4 s. As can be seen from Figure~\ref{fig:Histogram_by_Attacks_train}, the duration of bonafide audios seems to follow a uniform distribution, ranging from 10 to 20 s. In contrast, the duration of the different attack categories approximates a skewed normal distribution with a mean close to 11 s. This bias in audio length distributions may be leveraged by detection models. 
%when classifying utterances during training.
%The longer average duration in the training dataset may lead to richer feature representations, allowing the model to capture more information, but could also pose a risk of overfitting, as the model may become too attuned to the specific characteristics of the training dataset. 

\noindent
{\bf Development:} In the development dataset, however, the audios are shorter, with an average duration of 7.08 s and a standard deviation of 1.84 s. As depicted in Figure~\ref{fig:Histogram_by_Attacks_dev}, the distributions also differ, with most classes following a distribution similar to a uniform one ranging from 3 to 10 s. Exceptions include attack A13, which lacks a clear distribution pattern, and attack A11, which contains some audios with a duration less than 3 s.
%These variations in audio duration could be crucial for enhancing model accuracy and performance, particularly in detecting anomalies in attack patterns.

%The significant difference in average durations between the training and development datasets could introduce challenges in model generalization.

\subsection{Delays}

%Since a significant bias was discovered in the ASVspoof 2019 database, we also analyzed the utterance delay\footnote{In this work, delay is defined as the amount of time between the beginning of an audio file and the speech onset.} distributions of the ASVspoof5 datasets. 
Since initial silences were found to introduce a bias in the ASVspoof 2019 database~\cite{Muller21}, we also analyzed the utterance delay\footnote{In this work, delay is defined as the amount of time between the beginning of an audio file and the speech onset.} distributions in the ASVspoof5 datasets.
We employed a voice activity detector to spot the onset of speech and, subsequently, calculate the delay on an utterance basis.

\begin{figure}[t]
\centering
\includegraphics[width=0.49\textwidth]{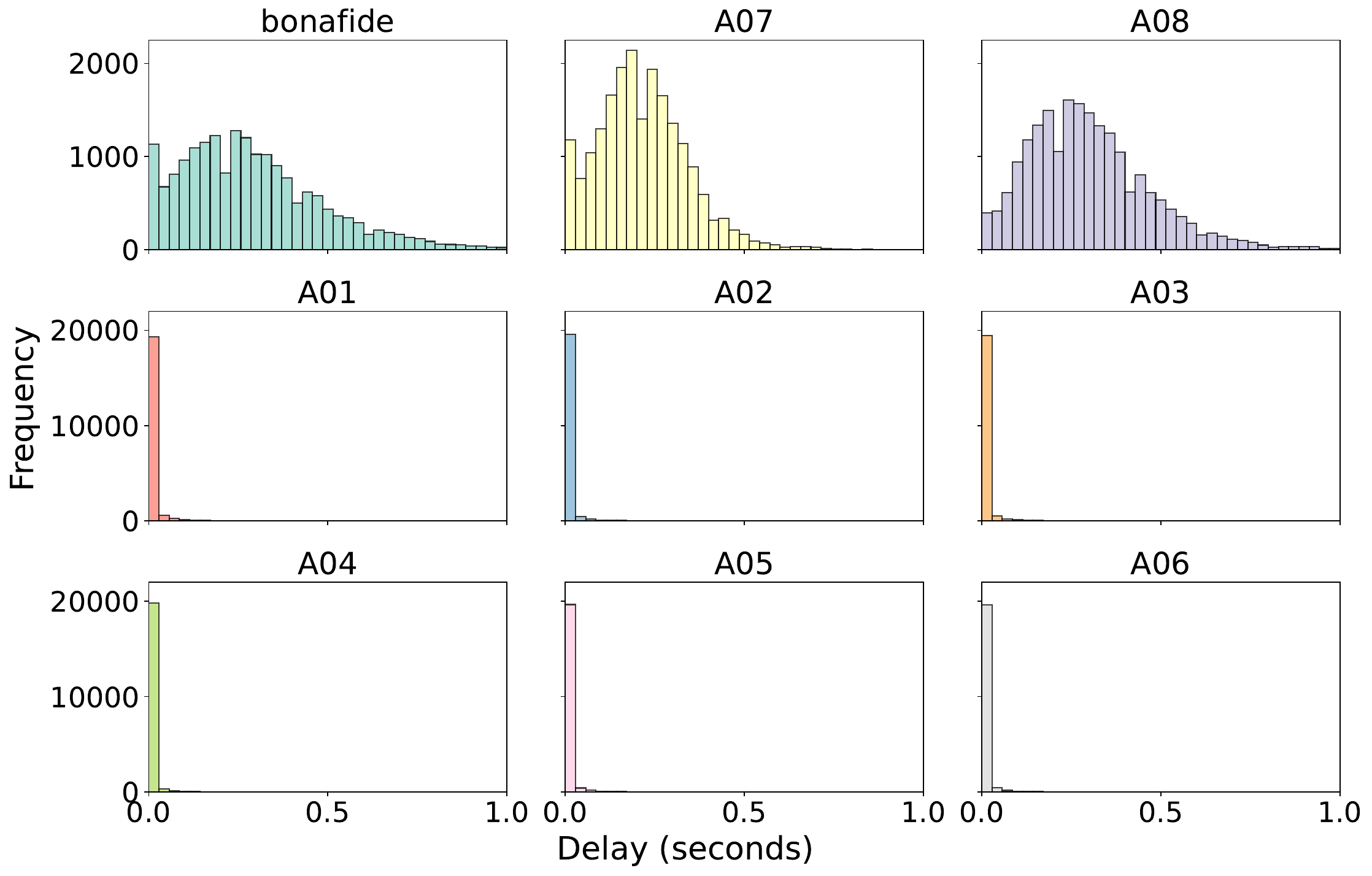}
%\caption{Training set utterance delay histograms.}\label{fig:Histogram_delays_train}
\caption{\it Histograms of utterance delay from the training set.}\label{fig:Histogram_delays_train}
%\vspace{0.35cm}
\end{figure}

\noindent
{\bf Training:} In the training dataset, the vast majority of audios appear to have no delay. However, it is important to note that those with delays are either bonafide samples or attack types A07 and A08, as Figure~\ref{fig:Histogram_delays_train} shows. This bias could potentially be exploited by a model to classify the audios, leading to overfit the training dataset.

\begin{figure}[t]
\includegraphics[width=0.49\textwidth]{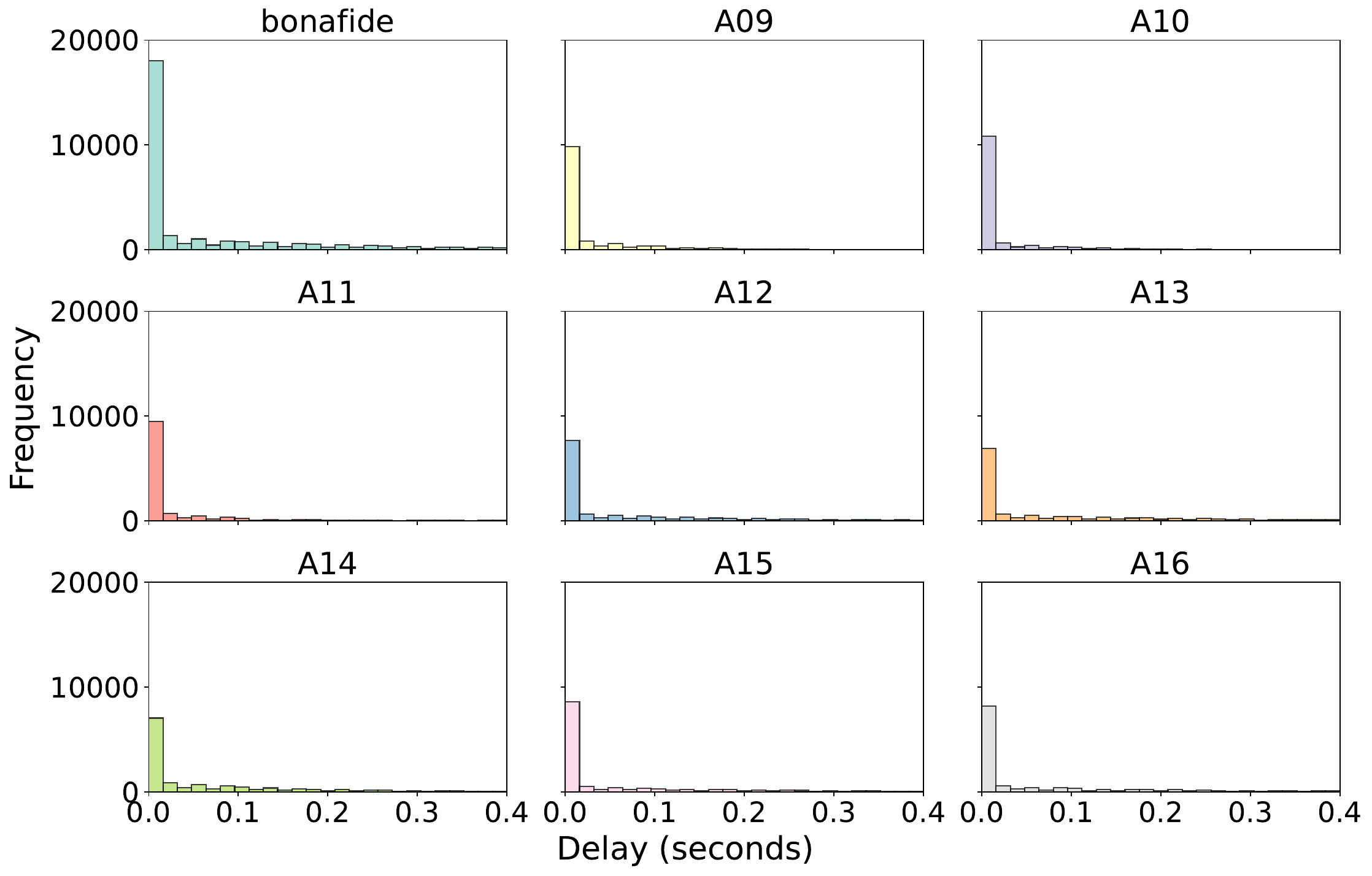}
%\caption{Development set utterance delay histograms.}\label{fig:Histogram_delays_dev}
\caption{\it Histograms of utterance delay from the development set.}\label{fig:Histogram_delays_dev}
\end{figure}

\noindent
{\bf Development:} In the development set, delays are almost negligible, with a short duration, and do not appear to be associated with any particular class. This contrasts with the case of the training set, where trimming or similar approaches could be advised.
%This suggests that delay is not a significant feature in distinguishing between classes within the development dataset, reducing the risk of overfitting the model based on delay characteristics alone.

\subsection{Speech quality}

\begin{figure}[t]
\centering
\includegraphics[width=0.49\textwidth]{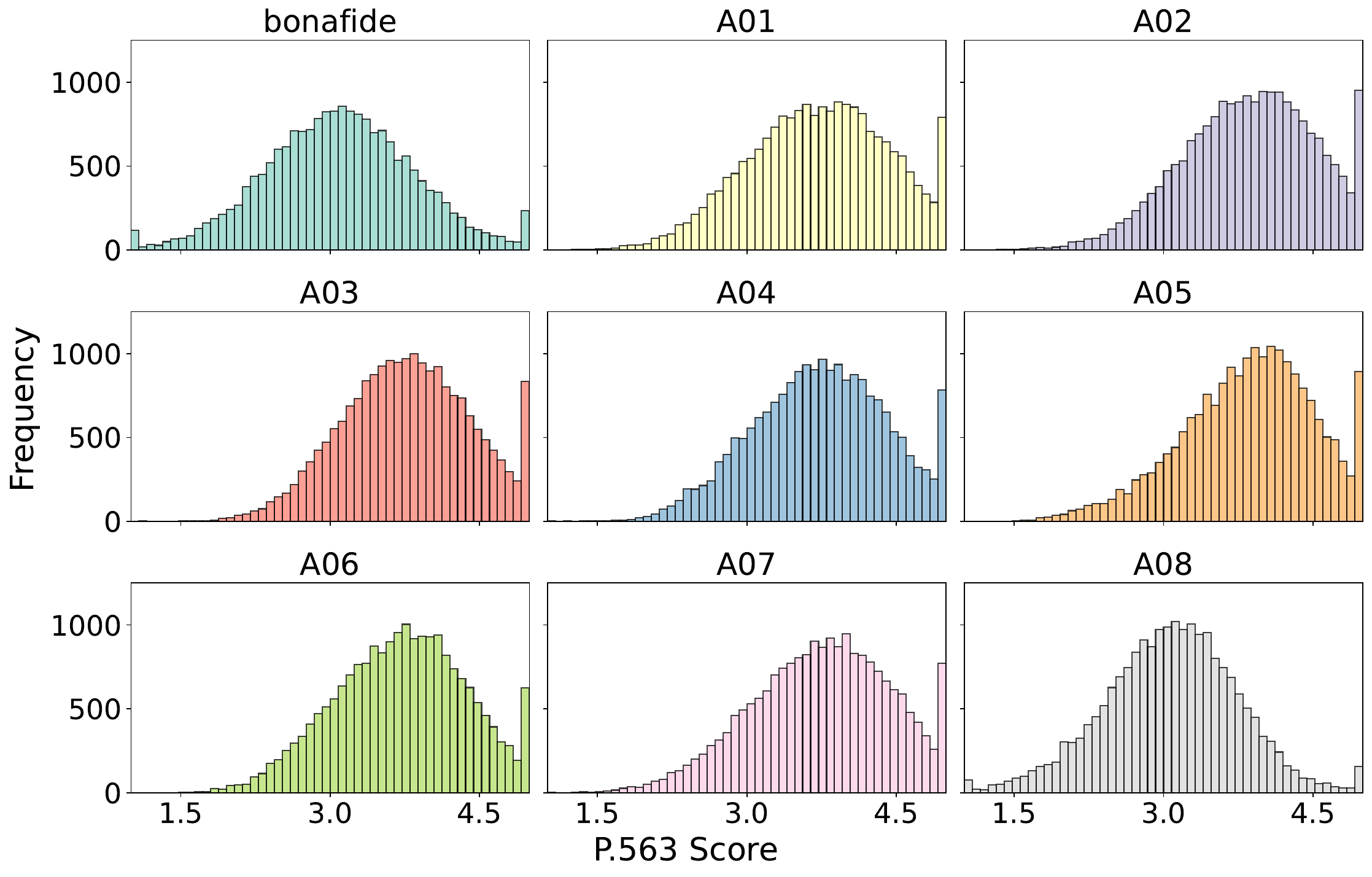}
%\caption{Training utterance P.563 score histograms for the different attack types.}\label{fig:Histogram_p563_train}
\caption{\it Histograms of P.563 scores for training utterances across different attack types.}\label{fig:Histogram_p563_train}
%\vspace{0.35cm}
\end{figure}

\begin{figure}[t]
\includegraphics[width=0.49\textwidth]{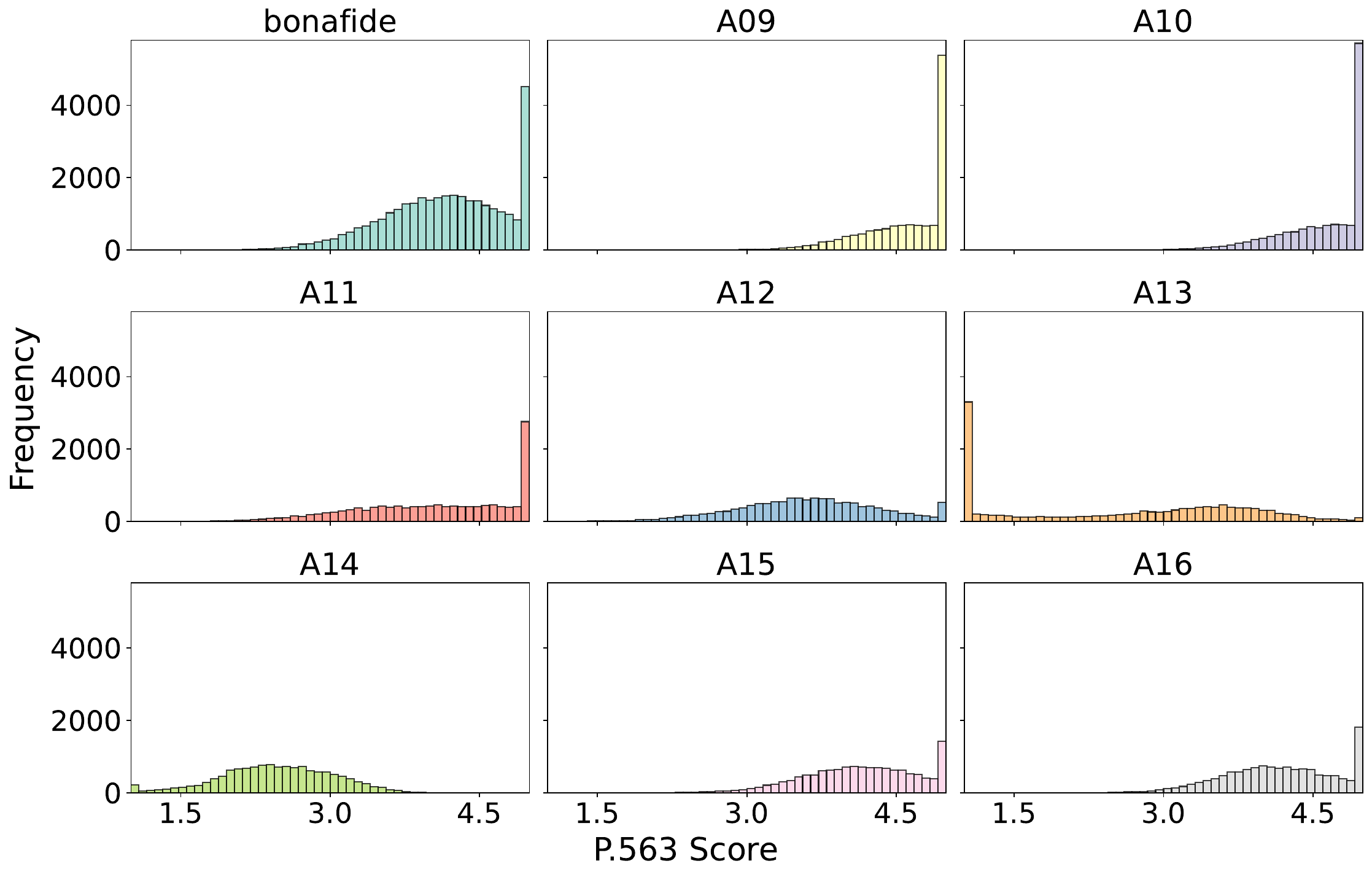}
%\caption{Development utterance P.563 score histograms for the different attack types.}\label{fig:Histogram_p563_dev}
\caption{\it Histograms of P.563 scores for development utterances across different attack types.}\label{fig:Histogram_p563_dev}
\end{figure}

\begin{figure}[t]
\centering
\includegraphics[width=0.49\textwidth]{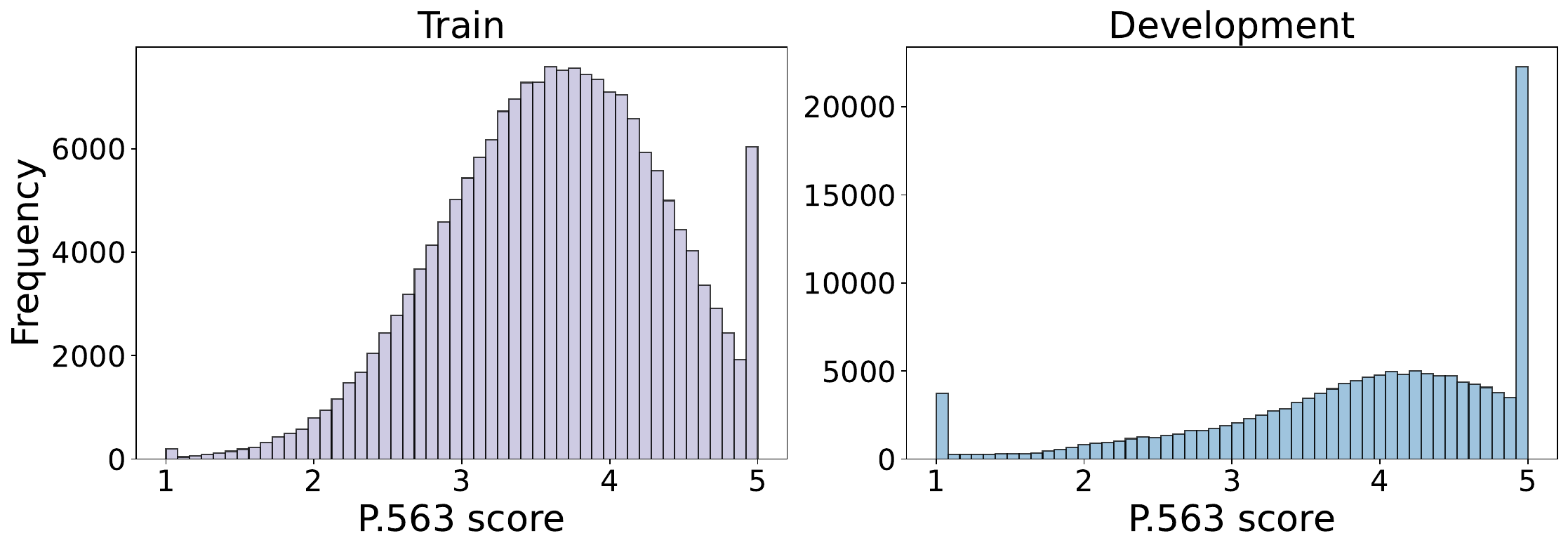}
\caption{\it Training and development utterance P.563 score histograms.}\label{fig:histogramas_generales_p563}
\end{figure}

Finally, in addition to the previous analyses, we conducted a speech quality evaluation on the training and validation datasets by means of the non-intrusive ITU-T standard P.563~\cite{ITU-P563} as an objective perceptual quality metric.

\noindent
{\bf Training:} On the training data, the speech quality scores seem to follow a quasi-normal distribution, with some clustering at the maximum value. The bonafide class has a slightly lower mean compared to most attack classes, as shown in Figure~\ref{fig:Histogram_p563_train}. Overall, the data show a diverse range of speech quality scores.

\noindent
{\bf Development:} In contrast to the above, the validation set appears to be less varied, exhibiting a more limited range and a tendency for values to cluster at the maximum. As illustrated by Figure~\ref{fig:Histogram_p563_dev}, there are two attacks, A13 and A14, with significantly lower quality metric values than the others. This pronounced difference may make these attacks easier to be distinguished from the other ones.

The disparity between the whole training and validation datasets is clearly evident from Figure~\ref{fig:histogramas_generales_p563}.
Although there is some clustering at the maximum value for the training data, the overall distribution is more varied compared to that of the validation dataset. Furthermore, the validation dataset tends to accumulate more heavily at the maximum and minimum values, indicating less diversity in speech quality.

\section{Closed-condition system}
\label{sec:closed}

Our team participated in the closed condition with a novel single system which, unfortunately, did not reach our expectations. Due to this, the system was only evaluated on Track 1. For completeness sake, the system is briefly described below.

\subsection{System description}

\noindent
\textbf{DNN model:} Our proposal is derived from the deep complex convolutional recurrent network (DCCRN) first described in~\cite{Hu20} for speech enhancement tasks.
This network consists of a causal convolutional encoder-decoder architecture with LSTM layers between the encoder and the decoder, so that the temporal dependencies can be modeled.
We chose this network because of its ability to process full spectra (i.e., both magnitude and phase), as the DCCRN is essentially an extension of a CRN that performs joint complex-value computation instead of considering two isolated real and imaginary parts.
For deepfake detection, we have removed the decoder part of the architecture.
Thus, the last hidden state of the deepest LSTM layer is used to compute an embedding from the received input, which is finally mapped into classes by a softmax layer.

%Table~\ref{tab:dccrn_arch} summarizes the structure of our derived network.

%obtained via STFT~\cite{Alteris07}, defined as:
%\begin{eqnarray}
%   X(l,k)=\sum_{n=0}^{M-1}  x(lH +n) w(n) e^{-j2\pi\frac{kn}{M}},
%\end{eqnarray}
%where $k$ is the frequency index (representing bins of $\frac{2\pi}{M}$ radians per sample), $l$ is the frame index, $x(n)$ the is input signal, and $w(n)$ is the analysis window.
%As can be noted, $x(n)$ is segmented in overlapped frames of length $M$ and shift $H$.
%In our proposal, we considered a square root Hann window for $w(n)$, and 512 and 128 samples for $M$ and $H$ ($32$ and $8$~ms at $16$ kHz), respectively.
\noindent
\textbf{Input data:} %according to the rules, only utterances from the training dataset were used for training. 
We set a fixed input of 96,000 samples ($6$ s), so that the network does not exploit any length bias (see Section~\ref{sec:asvspoof5}).
Similarly, audio excerpts are extracted from the middle of the utterance during validation and testing, and from a random position during training, in order to prevent the network from learning of the initial delays. 
%If the utterance length is shorter than 96,000 samples, samples are repeated until the previous condition is meet. 
A complex-spectrum representation is then obtained via STFT~\cite{Alteris07} with a square-root Hann window of $512$ samples shifted by $128$ ($32$ and $8$~ms at $16$ kHz).
Thus, input tensors are of size $(C \times F \times T)$, with $C=2$ channels (real and imaginary parts), $F=256$ bins of frequency (DC component is removed) and $T=750$ frames.

%We use the last hidden state from the last LSTM layer to compute an embedding from the received input.
%To this end, a linear layer followed by batch normalization is used. Then, another linear layer is applied to map embeddings into classes.
\noindent
\textbf{Training setup:}
Network optimization is performed through the ADAM algorithm~\cite{kingma2014adam} with a batch size of 64 utterances and a learning rate of $3\cdot10^{-4}$.
We use the weighted cross entropy (WCE)~\cite{lin2017focal} as a loss function during training.
Despite the excellent convergence on the training dataset, WCE scores were very pathological during validation, diverging after the first epoch. 
%Divergent WCE values were observed after the first epoch on the validation dataset (not only for this network but other tested architectures), and this issue could not be resolved by adjusting the learning rate or changing the training loss function (we also tested multi-class training as well as one-class softmax loss). 
As this issue could not be resolved by adjusting the learning rate or changing the training loss function, we trained the model for 100 epochs and selected the one that achieved the best EER on the validation dataset (provided it was not a transient spike). 

\subsection{Challenge results}
As mentioned above, model performance was not satisfactory, mainly due to the validation issues found and the apparent inability of the network to generalize.
Table~\ref{table2} summarizes the results obtained during the progress and evaluation phases of the challenge.
As can be observed, our proposal improves the AASIST~\cite{jung2022aasist} baseline, provided by the organizers~\cite{Wang2024_ASVspoof5}, in terms of minDCF ($0.6598$ vs. $0.7110$), but EER is marginally the same ($28.41$\% vs. $29.12$\%).

\begin{table}[!t]
\caption{\label{table2} {\it Track 1 closed-condition results provided by the DCCRN model.}}
\vspace{2mm}
\centerline{
\begin{tabular}{ccccc}
\toprule
\textbf{Phase} & minDCF & actDCF & C$_{\mbox{\scriptsize llr}}$ \cite{van2007introduction} & EER (\%)\\
\midrule 
\textit{Progress} & 0.4591 & 1.0000 & 1.0426 & 18.63 \\
\textit{Evaluation} & 0.6598 & 1.0000 & 1.1159 & 28.41 \\
\bottomrule
\end{tabular}}
\end{table}

\section{Open-condition systems}
\label{sec:open}

\subsection{System description}
Our team also participated in the open condition with an ensemble system that yielded significantly better results. This system demonstrated improved performance and robustness in both Track 1 and 2. Below, we provide a description of this successful system.

\subsubsection{Track 1: Speech deepfake detection}

%We now describe our proposed system for the Track 1 open condition, including the involved DNN models, training data, data augmentation techniques, calibration, and ensemble fusion.

\noindent
\textbf{DNN models:} Our approach is based on pre-trained self-supervised learning (SSL) speech models as feature extractors to obtain robust deep embeddings. We considered two different SSL models pre-trained on the LibriSpeech corpus: Wav2Vec2-Large (W2V2) \cite{Baevski20} and WavLM-Base (WavLM) \cite{Chen22}. The SSL deep embeddings are then converted to a final spoof score through a downstream model fine-tuned on the target data. This downstream model is different depending on the corresponding SSL upstream considered. 
%This is because padding masks are used on the WavLM and surrogate downstream during training, which are not considered in the W2V2 model following model recommendations. 
This is because padding masks are used in WavLM (and surrogate downstream) during training, but not in the W2V2 model, following model recommendations.
Therefore, the classifier for our winner system in ADD 2022 Track 1 \cite{martin2022vicomtech} is combined with the W2V2 feature extractor. On the other hand, for WavLM upstream, we chose the NN-ASP classifier from our recent paper \cite{martin2024exploring}, which takes into account the padding mask. Nevertheless, the structure of those downstreams is similar, following weighted sum of the Transformer layers, per-frame non-linear transformations, attentive statistical pooling, and cosine scoring. These downstreams are trained by minimizing the one-class softmax loss \cite{zhang2021one}.

\noindent
\textbf{Train and development data:} The main corpus for training is the ASVspoof5 dataset previously described. To boost deepfake detection and robustness against different attacks, we extended this corpus with external databases. Thus, we included the training and development data from the 2019 ASVspoof Challenge \cite{nautsch2021asvspoof}, based on the VCTK database. Moreover, we extended this dataset by aggregating additional vocoded data as described in \cite{wang2023spoofed}. We used the \texttt{Voc.v4} partition, which considers four additional vocoders pre-trained on LibriSpeech and fine-tuned on ASVspoof 2019 bonafide speech. The train and development sets from the in- and out-domain data are not mixed, i.e., original train data are only augmented with the train sets of the out-domain datasets (similar with development partitions).

\noindent
\textbf{Data augmentation:} We applied on-the-fly data augmentation techniques during the training of our systems. First, we trimmed the leading and trailing silences of the signals to avoid exploiting potential misleading artifacts from the databases, especially from the ASVspoof 2019 dataset. Then, to emulate the effect of different codec systems on the clean speech signals, we applied RawBoost data augmentation \cite{tak2022rawboost}. In this case, we considered the full configuration of RawBoost including three kinds of distortions: linear and non-linear convolutive noise, impulsive signal-dependent additive noise, and stationary signal-independent additive noise.

\noindent
\textbf{Training setup:} The models are fine-tuned on the corresponding training data using the ADAM optimizer \cite{kingma2014adam} with a learning rate of $3\cdot 10^{-4}$. The effective batch size is 64 utterances (8 utterances mini-batch and 8 steps for gradient accumulation). On the other hand, the development data are only considered for best model selection and early-stopping.

\noindent
\textbf{Calibration:} The output cosine scores of the detection systems are not well-calibrated LLRs, making those scores sub-optimal for proper Bayesian decisions \cite{van2007introduction}. To reduce the calibration loss, we trained an additional calibration backend using the ASVspoof5 development set. The calibrator is only restricted to be a monotonic rising function converting the raw scores to LLRs calibrated on the development set. Apart from the well-known logistic regression (LogReg), defined as a linear function for LLR scores, we considered the beta calibration \cite{kull2017beta}, which assumes beta distributions for the scores. 
%Therefore, this calibration method seems to be more appropriate for cosine scores limited in range. 
Beta calibration seems better suited for cosine scores within a specific range.
We applied the univariate version of the calibration function, defined as 
$L(s') = a \cdot \log{\frac{s'}{1-s'}} + c$, where $s'$ are the scaled scores in the range $[0,1]$, while $a \geq 0$ and $c \in \mathbb{R}$ are the function parameters to be fitted. It should be noticed that this calibrator can be trained as a LogReg by first converting the scaled scores through the logarithmic function.

\noindent
\textbf{Ensemble system:} Finally, we also explored the ensemble of our best-performing detection systems through late score fusion. To this end, the output scores from both the W2V2 and WavLM subsystems are combined by a linear weighted sum. We evaluated the fusion of both raw scores and calibrated LLRs.

\begin{table*}[!t] 
\caption{\it Track 1 open-condition results for the different systems proposed during the progress phase.}
\vspace{2mm}
\label{table:t1open_progress}
\centering
\begin{tabular}{ccc|cccc}
\toprule
\multirow{2}{*}{\textbf{Model}} & \multirow{2}{*}{\textbf{Data}} & \multirow{2}{*}{\textbf{Calibrator}} & \multicolumn{4}{c}{\textbf{Result}} \\
 {} & {} & {} & minDCF & actDCF & C$_{\mbox{\scriptsize llr}}$ & EER (\%) \\ 
\midrule
\multirow{8}{*}{W2V2} & asv19 & -- & 0.3419 & 0.9818 & 0.7119 & 11.79 \\
{} & asv19voc & -- & 0.2513 & 0.9983 & 0.7220 & 8.75 \\
{} & asv5 & -- & 0.0550 & 0.2679 & 0.5671 & 2.02 \\
{} & asv5 & LogReg & 0.0550 & \textbf{0.0563} & 0.2000 & 2.02 \\
{} & asv5 & Beta & 0.0550 & 0.0762 & 0.1440 & 2.02 \\
{} & asv5+19voc& -- & 0.0354 & 0.5647 & 0.5720 & 1.23 \\
{} & asv5+19voc & LogReg & 0.0354 & 0.0699 & 0.1711 & 1.23 \\
{} & asv5+19voc & Beta & 0.0354 & 0.0893 & 0.1254 & 1.23 \\
\midrule
\multirow{3}{*}{WavLM} & asv5 & -- & 0.0820 & 0.2271 & 0.5597 & 3.15 \\
{} & asv5+19voc & -- & 0.0319 & 0.0661 & 0.5048 & 1.16 \\
{} & asv5+19voc & Beta & 0.0319 & 0.1423 & 0.2092 & 1.16 \\
\midrule
\multirow{2}{*}{Ensemble} & asv5+19voc & -- & \textbf{0.0186} & 0.2385 & 0.5368 & \textbf{0.65} \\
{} & asv5+19voc & Beta & \textbf{0.0186} & 0.0843 & \textbf{0.1133} & \textbf{0.65} \\
\bottomrule
\end{tabular}
\end{table*}

\subsubsection{Track 2: Spoofing-aware ASV}

For the SASV open track, we proposed a straightforward combination of ASV and CM subsystems by score fusion. This allowed us to quickly deploy and evaluate an SASV system based on a fixed ASV model and our best CM system for Track 1.

For the ASV subsystem, we considered the pre-trained TitaNet-Large model included in the Nvidia NeMo toolkit \cite{TitaNet}. This architecture is based on a convolutional network with squeeze-and-excitation layers and channel attention to extract a speaker embedding from an utterance. This network is trained with multiple ASV corpora, including VoxCeleb 1 and 2, LibriSpeech, and telephonic data (NIST SRE 04--08, Fisher and Switchboard). The embedding for the target speaker in each trial is obtained by averaging the embeddings from the different enrollment utterances of the speaker. Two different scoring backends are evaluated: cosine scoring, and a probabilistic linear discriminant analysis (PLDA) model trained on the ASVspoof5 bonafide training data. Moreover, the output scores are also calibrated on the ASVspoof5 development set (considering target and non-target trials only) by means of LogReg. This calibration is especially important for proper integration with the calibrated CM scores.

Finally, for score fusion, we compared two different approaches. The first one is simply a linear weighted sum of the scores. A better procedure is proposed in \cite{wangRevisiting2024} based on a non-linear fusion of LLR scores. To this end, a negative LogSumExp (LSE) function (smooth maximum) is applied to the negative LLR scores, yielding the final SASV LLR scores. Furthermore, a different weight is considered for the scores during the sum, which is adjusted by a grid search on the development data. 
%Note that, in contrast to \cite{wangRevisiting2024}, in our approach we do not fit Gaussian distributions to the raw score vectors. Instead, we directly use the LLRs from the individual subsystems, avoiding possible overfitting to the train/development datasets.
Note that, in contrast to \cite{wangRevisiting2024}, our approach does not fit Gaussian distributions to the raw score vectors. Instead, we directly use the LLRs from the individual subsystems, thereby avoiding potential overfitting to the training/development datasets.

\subsection{Challenge results}

\subsubsection{Track 1}

We first evaluated the different configurations for our proposed approach on the progress subset. Table~\ref{table:t1open_progress} shows the results achieved for different combinations of DNN models, training data and calibration method. With respect to the data, we can observe that the combination of ASVspoof5 and ASVspoof 2019 generally yields the best results, and using additional vocoding data is beneficial for generalization purposes (comparing asv19 and asv19voc experiments). Calibration also helps improve secondary metrics, with beta calibration giving better results through different operation points (lower C$_{\mbox{\scriptsize llr}}$ values). WavLM models produce better results than W2V2 when the aggregated data are used for training, and the model ensemble fusion yields the best performance. Thus, we considered this ensemble model with beta calibration as our final system for the subsequent evaluation phase.

\begin{table}[!t]
\caption{\it Evaluation phase results for our submitted system in Track 1 open condition.}
\vspace{2mm}
\label{table:t1open_eval}
\centering
\begin{tabular}{cccc}
\toprule
minDCF & actDCF & C$_{\mbox{\scriptsize llr}}$ & EER (\%) \\
\midrule
0.1348 & 0.2170 & 0.3096 & 5.02 \\
\bottomrule
\end{tabular}
\end{table}

\begin{table*}[th]
%\caption{EER (\%) results for the Track 1 open condition for our submitted system broken down by spoofing attack and codec condition.}
\caption{\it Track 1 EER (\%) results achieved by our open-condition system, broken down by spoofing attack and codec condition.}
\vspace{2mm}
\label{table:t1open_cond}
\centering
\begin{tabular}{c|cccccccccccc|c}
\toprule
         &    --     & C01  & C02  & C03  & C04  & C05  & C06  & C07  & C08  & C09 & C10 & C11 &  Pooled \\ 
\midrule
 A17   &   0.05  &   0.20  &   0.17  &   0.37  &   0.50  &   0.05  &   0.15  &   1.00  &   1.30  &   0.43 &   0.79  &   0.12  &   0.72   \\ 
 A18   &   0.15  &   1.47  &   1.38  &   1.81  &   3.35  &   0.37  &   0.68  &   4.43  &   2.85  &   2.70 &   5.13  &   0.38  &   2.58   \\ 
 A19   &   0.10  &   0.66  &   0.51  &   0.91  &   2.13  &   0.14  &   0.27  &   2.74  &   1.35  &   0.49 &   1.30  &   0.14  &   1.38   \\ 
 A20   &   0.15  &   1.06  &   1.05  &   1.76  &   3.36  &   0.24  &   0.64  &   5.25  &   2.77  &   1.49 &   4.32  &   0.18  &   2.16   \\ 
 A21   &   0.01  &   0.29  &   0.17  &   0.41  &   0.92  &   0.04  &   0.09  &   1.22  &   1.13  &   0.24 &   0.71  &   0.05  &   0.64  \\ 
 A22   &   0.10  &   0.84  &   0.67  &   1.65  &   3.15  &   0.25  &   0.44  &   4.55  &   2.18  &   1.55 &   2.92  &   0.21  &   1.84   \\ 
 A23   &   0.13  &   1.01  &   1.16  &   1.94  &   3.48  &   0.29  &   0.59  &   5.02  &   2.34  &   1.79 &   3.79  &   0.21  &   2.22   \\ 
 A24   &   0.21  &   1.14  &   1.05  &   2.48  &   3.61  &   0.33  &   0.58  &   5.15  &   2.59  &   1.31 &   3.63  &   0.29  &   2.39   \\ 
 A25   &   0.05  &   0.30  &   0.33  &   0.62  &   1.66  &   0.12  &   0.15  &   2.54  &   1.59  &   0.49 &   1.54  &   0.05  &   1.02   \\ 
 A26    &   0.05  &   0.42  &   0.34  &   1.11  &   2.12  &   0.09  &   0.19  &   2.67  &   1.76  &   0.68 &   1.96  &   0.12  &   1.26 \\ 
 A27   &   0.41  &   3.36  &   3.97  &   6.38  &   8.13  &   0.84  &   1.90  &  11.02  &   7.32  &   6.22  &  12.46  &   0.63  &   5.45   \\ 
 A28   &   4.99  &  13.57  &  12.09  &  13.90  &  20.34  &   6.57  &   9.18  &  24.21  &  22.15  &  18.15 &  26.66  &  10.60  &  16.10   \\ 
 A29   &   0.15  &   0.50  &   0.42  &   0.58  &   0.54  &   0.33  &   0.34  &   0.58  &   1.00  &   0.53 &   0.50  &   0.42  &   0.63   \\ 
 A30   &   0.37  &   2.99  &   2.70  &   3.91  &   6.07  &   0.75  &   1.42  &   8.18  &   5.52  &   4.44 &   9.16  &   0.50  &   4.47   \\ 
 A31   &   0.79  &   4.40  &   4.61  &   6.63  &  10.21  &   1.27  &   1.91  &  12.86  &   6.97  &   6.32 &  10.73  &   0.75  &   5.87   \\ 
 A32   &   0.29  &   2.94  &  3.23  &   5.47  &   7.67  &   0.68  &   1.64  &   9.70  &   7.22  &   5.78 &  12.46  &   0.69  &    5.02   \\
\midrule
 Pooled &   1.17  &   3.75  &   3.46  &   4.67  &   6.60  &   1.70  &   2.27  &   8.28  &   6.24  &   4.95 &   8.58  &   2.29  &   5.02   \\ 
\bottomrule
\end{tabular}
\end{table*}

\begin{table*}[!t] 
\vspace{-1.5mm}
\caption{\it Track 2 open-condition results for the different systems evaluated during the progress phase.}
\vspace{2mm}
\label{table:t2open_progress}
\centering
\begin{tabular}{cccc|ccc}
\toprule
\multirow{2}{*}{\textbf{CM}} & \multirow{2}{*}{\textbf{ASV Backend}} & \multirow{2}{*}{\textbf{Fusion}} & \multirow{2}{*}{\textbf{Calib.}} & \multicolumn{3}{c}{\textbf{Result}} \\
 {} & {} & {} & {} & min a-DCF \cite{shim24adcf} & min t-DCF \cite{kinnunen2020tandem} & t-EER (\%) \cite{kinnunen2024teer} \\ 
\midrule
WavLM & Cosine & Linear & $\times$ & 0.1700 & 0.1240 & 4.08 \\
\multirow{5}{*}{Ensemble} & Cosine & Linear & $\times$ & 0.1436 & 0.1102 & 3.96 \\
{} & Cosine & LSE ($p=0.5$) & \checkmark & 0.0708 & \textbf{0.1093} & 3.97 \\
{} & Cosine & LSE ($p=0.7$) & \checkmark & \textbf{0.0661} & \textbf{0.1093} & 3.97 \\
{} & PLDA & LSE ($p=0.5$) & \checkmark & 0.0752 & \textbf{0.1093} & \textbf{3.83} \\
{} & PLDA & LSE ($p=0.7$) & \checkmark & 0.0682 & \textbf{0.1093} & \textbf{3.83} \\
\bottomrule
\end{tabular}
\end{table*}

\begin{table}[th]
\vspace{-1.5mm}
\caption{\it Evaluation phase results for our submitted system in Track 2 open condition.}
\vspace{2mm}
\label{table:t2open_eval}
\centering
\begin{tabular}{cccc}
\toprule
min a-DCF & min t-DCF & t-EER (\%) \\
\midrule
0.1295 & 0.4372 & 5.43 \\
\bottomrule
\end{tabular}
\end{table}

Our final results for the evaluation phase are shown in Table~\ref{table:t1open_eval}. We achieve competitive results with a 0.1348 minDCF and an EER of 5.02\% while also keeping good performance in terms of the other two metrics that measure both discrimination and calibration capabilities. To disentangle these results, we show EER values broken down by spoofing attack and codec in Table~\ref{table:t1open_cond}. We choose EER instead of the primary metric minDCF since EER makes the comparison easier. That being said, both metrics are directly related, and similar trends can be observed. Results in terms of the attack type reveal that our system is mainly negatively affected by A28 (16.10\% EER). This spoof attack corresponds to a pre-trained YourTTS model \cite{casanova2022yourtts}. It is interesting to note that a similar attack using YourTTS is also included in the development set, where this performance degradation was not observed, which can be due to different configurations or modifications. Further investigations will be needed to comprehend this difference. We can also see that attacks A27 and A30-32 yield a higher EER ($\sim$5\%) in comparison with other spoof attacks. The common feature of those spoofing systems is that they also include the Malacopula adversarial attack \cite{todisco2024malacopula}. Although our approach generally behaves properly against adversarial attacks, further countermeasures should be taken into account to improve the results under A27 and A30-32. On the other hand, results across codec conditions show that the good performance in clean conditions (1.17\% EER) is severally degraded when using codecs C07 (MP3+Encodec) and C10 (Speex 8 kHz), and moderately degraded by C04 (Encodec) and C08 (Opus 8 kHz). In general, our approach is more affected under narrowband conditions (8 kHz) and neural audio compression (Encodec \cite{defossezhigh}). Although the RawBoost data augmentation helps better generalize across codec conditions, it cannot completely cover these two scenarios, probably requiring additional augmentation techniques that can cope with the new degradations and artifacts produced by these channel codecs. Nonetheless, it can be observed that the performance of our approach is generally robust and competitive across a broad set of codecs and spoofing attacks.

\subsubsection{Track 2}

Table~\ref{table:t2open_progress} depicts our results for the Track 2 open condition during the progress phase. Due to the limited amount of trials, we mainly considered our best fusion configurations evaluated on the development set, which are based on LSE score fusion from calibrated systems. This table also includes linear fusion using cosine scores with both WavLM and ensemble CM as baselines. As can be observed, using LSE fusion with calibrated LLRs yields better performance than a simple linear fusion, especially when considering the minimum a-DCF metric (a-DCF from now on). Adjusting the weight between CM and ASV scores can also improve the results. In our case, higher weights $p$ for the ASV scores produced, in general, better performance. Finally, a comparison between ASV backends reveals that similar performances are achieved, with the cosine (PLDA) scoring outperforming in terms of a-DCF (t-EER). This demonstrates that the pre-trained ASV embedding extractor is competitive enough to obtain good verification performance. Thus, we selected cosine scoring as our ASV backend, and combined the calibrated LLR scores from CM and ASV subsystems using the LSE score fusion.

Finally, we present our results for the evaluation phase in Table~\ref{table:t2open_eval}. We achieved a competitive a-DCF of 0.1295 with our proposed system, as well as a strong performance in terms of other tandem-related metrics. Regarding results per attack type and codec condition, we observed similar trends to those from our Track 1 results. The most challenging attack was A28 (0.451 a-DCF), followed by the systems using the Malacopula adversarial attack (results within the range $[0.111, 0.153]$). Moreover, the most challenging codecs were C08 and C10, with results close to 0.195 a-DCF, especially compared to clean conditions (0.055 a-DCF). Nevertheless, we achieved a robust and straightforward score fusion approach based on reliable ASV and CM subsystems, resulting in competitive challenge results.

\section{Conclusion}
\label{sec:conclusion}

In this paper, we have presented the Vicomtech-UGR systems submitted to the ASVspoof5 Challenge. After facing difficulties in developing CM systems for the Track 1 closed condition, we achieved a robust ensemble system with competitive performance in the open condition. This was due to leveraging self-supervised models, and augmented training data from previous challenges and novel vocoders. For the SASV system of Track 2, we have combined our ensemble CM system with a pre-trained ASV model via a straightforward non-linear score fusion. For both tracks, calibration has been a key aspect to provide meaningful LLR scores, especially during the integration of ASV and CM subsystems. As future work, we will analyze the robustness of our speech deepfake detection approach against state-of-the-art speech synthesis models, and the development of advanced data augmentation techniques covering additional codecs, narrowband channels, and adversarial attacks.

%This template can be found on the conference website
%$<$https://www.asvspoof.org/workshop5$>$. \cite{aluisio2001learn}

\section{Acknowledgements}
\label{sec:ack}
This work is part of the project PID2022-138711OB-I00 funded by MICIU/AEI/10.13039/501100011033 and by ERDF/EU, and the FPI grant PRE2022-000363. Also, this work has been supported in part by the European Union’s Horizon Europe research and innovation programme in the context of project EITHOS under Grant Agreement No. 101073928.

\newpage

\bibliographystyle{IEEEbib}
\bibliography{ASVspoof_BibEntries}

\end{document}